\documentclass[10pt,conference]{IEEEtran}
\IEEEoverridecommandlockouts

\usepackage{cite}
\usepackage{hyperref}
\usepackage{orcidlink}
\usepackage{comment}
\usepackage{amsmath,amssymb,amsfonts}
\usepackage{algorithmic}
\usepackage{graphicx}
\usepackage{textcomp}
\usepackage{xcolor}
\usepackage{soul}
\usepackage{acronym}
\usepackage{multicol}
\usepackage{multirow}
\usepackage{subfigure}

\def\BibTeX{{\rm B\kern-.05em{\sc i\kern-.025em b}\kern-.08em
    T\kern-.1667em\lower.7ex\hbox{E}\kern-.125emX}}

\begin{document}

\title{Batteryless BLE and Light-based IoT Sensor Nodes for Reliable Environmental Sensing\vspace{-0.1em}}

\author{\IEEEauthorblockN{Jimmy Fernandez Landivar$^{\mathrm{1}\diamond} $\orcidlink{0000-0002-4904-5256}, Khojiakbar Botirov$^{\mathrm{2}\diamond}$\orcidlink{0009-0004-6690-3365}, Hazem Sallouha$^{\mathrm{1}}$\orcidlink{0000-0002-1288-1023}, Marcos Katz$^{\mathrm{2}}$\orcidlink{0000-0001-9901-5023} and Sofie Pollin$^{\mathrm{1}}$\orcidlink{0000-0002-1470-2076}}
\IEEEauthorblockA{\textit{$^{\mathrm{1}}$Department of Electrical Engineering (ESAT) - WaveCoRE}, KU Leuven, 3000 Leuven, Belgium \\
\textit{$^{\mathrm{2}}$Centre for Wireless Communications}, University of Oulu, 90570 Oulu, Finland\\
\textit{$^{\diamond}$Authors contributed equally}\\ E-mails: \{jfernand, hazem.sallouha, sofie.pollin\}@esat.kuleuven.be, \{khojiakbar.botirov, marcos.katz\}@oulu.fi}\\

\vspace{-3em}

\thanks{The present work has received funding from the European Union's Horizon 2020 Marie Skłodowska Curie Innovative Training Network Greenedge (GA. No. 953775). The work of  Hazem  Sallouha was funded by the  Research Foundation – Flanders (FWO), Postdoctoral Fellowship No. 12ZE222N.}

}

\maketitle
\begin{abstract}

The sustainable design of Internet of Things (IoT) networks encompasses considerations related to energy efficiency and autonomy as well as considerations related to reliable communications, ensuring no energy is wasted on undelivered data.
Under these considerations, this work proposes the design and implementation of energy-efficient Bluetooth Low Energy (BLE) and Light-based IoT (LIoT) batteryless IoT sensor nodes powered by an indoor light Energy Harvesting Unit (EHU). Our design intends to integrate these nodes into a sensing network to improve its reliability by combining both technologies and taking advantage of their features. The nodes incorporate state-of-the-art components, such as low-power sensors and efficient System-on-Chips (SoCs). Moreover, we design a strategy for adaptive switching between active and sleep cycles as a function of the available energy, allowing the IoT nodes to continuously operate without batteries. Our results show that by adapting the duty cycle of the BLE and LIoT nodes depending on the environment's light intensity, we can ensure a continuous and reliable node operation. In particular, measurements show that our proposed BLE and LIoT node designs are able to communicate with an IoT gateway in a bidirectional way, every 19.3 and 624.6 seconds, respectively, in an energy-autonomous and reliable manner.

\end{abstract}


\section{Introduction}
Environmental sensing and monitoring rely on the dense deployment of sensors that are needed to ensure accurate data acquisition. To guarantee a reliable, energy and cost-efficient deployment, the focus is on implementing energy-autonomous devices, efficient low-power communication protocols, and adopting sustainable design and deployment approaches~\cite{sallouha2017ulora,9083805}. Therefore, cutting-edge technologies such as batteryless Internet of Things (IoT) devices aim to minimize hardware elements by employing organic electronic materials such as Printed Electronics (PE) while integrating highly efficient sensors and alternative environmentally friendly power sources like light or Radio-Frequency (RF) energy harvesting. This technology aims to establish sustainable IoT nodes following and spreading the ``Expose and Connect'' concept; whenever a node is exposed to an energy source, it connects to the network and sends information~\cite{9083805}. For instance, Bluetooth Low Energy (BLE) and Visible Light Communications (VLC) are promising technologies that support this concept while offering reliable communications.   

In particular, BLE is among the most suitable technologies for monitoring applications for many reasons, such as its ease of integration with multiple sensor devices, its cost-effectiveness in network deployment, and the advantage of being integrated into highly efficient System-on-Chips~(SoCs) with low energy consumption features. These reasons allow BLE to be used in short-range communications with minimal energy consumption through optimized scanning and advertising cycles~\cite{Santejudean2020}. However, in extensive IoT sensor network deployments, BLE and RF technologies may be subject to congestion and interference issues~\cite{9083805}. 
VLC has recently evolved as an IoT low-power technology capable of tackling the mentioned challenges of BLE networks by using the light spectrum for wireless data exchange.
A Light-based IoT (LIoT) node can exploit different frequencies of the visible and non-visible light spectrum for the uplink and downlink, such as Infrared (IR) and Visible Light (VL)~\cite{Botirov2023,s21238024}. Thus, implementing an IoT network with BLE and LIoT can enhance its reliability and node efficiency by leveraging both technologies' strengths without mutual interference\cite{Abuella2021}. In addition to the communication technologies, integrating low-power sensors and energy-efficient algorithms is essential to ensure reliable communications despite low-energy constraints~\cite{Santejudean2020,Botirov2023}.

Moreover, environmental sensing is critical to creating a controlled and safe environment in several use cases such as indoor laboratories, manufacturing plants, and office buildings~\cite{Santejudean2020,Liu2021}. Precisely, sensing parameters such as temperature, humidity, and Volatile Organic Compounds (VOCs) contribute to assessing and improving the infrastructure Indoor Air Quality~(IAQ)~\cite{Santejudean2020, Liu2021}. In these scenarios, the deployment of monitoring networks with reliable IoT sensor nodes implementing BLE and LIoT technologies can support these critical monitoring networks with energy-efficient sensing operations.   

Concerning this topic, this paper proposes the design of BLE and LIoT sensor nodes using state-of-the-art software and components, such as low-power sensors and SoCs. Our solution follows a sustainable approach by using PE and avoiding using batteries. These sensor nodes are intended to integrate into an indoor environmental sensing network, providing a reliable operation under constant illumination.

Hence, our main contributions are summarized as follows:
\begin{itemize}
\item We investigate the design of two batteryless IoT nodes, one based on BLE and another one on LIoT. The sensor nodes use state-of-the-art components, such as a sustainable Energy Harvesting Unit (EHU), low-power sensors, and energy-efficient communication technologies. 
\item Based on our design, a solution for the sensing, communication, and batteryless low-power operation is proposed. It includes a detailed analysis of the nodes' active and sleep cycles for low-power periodic operation based on experimental energy consumption measurements.
\item Finally, we conducted an eight-hour experimental evaluation to demonstrate the energy conservation of the nodes during active and sleep cycles and the stability of the implemented solution in a continuous operational period.
\end{itemize} 

The rest of this paper is organized as follows: Section~\ref{Sec_II} presents the related work. Section~\ref{Sec_III} describes the principal components of a batteryless IoT sensor node. Sections~\ref{Sec_IV} and~\ref{Sec_V} detail the energy consumption analysis and implementation of the BLE and LIoT batteryless sensor nodes, respectively. Section~\ref{Sec_VI} presents the experimental performance evaluation of our solution, and Section~\ref{Sec_VIII} concludes and outlines future work.

\section{Related work}
\label{Sec_II}
Energy-efficient IoT devices for sensing have been widely investigated. For instance, \cite{Santejudean2020} designed a device using BLE, powered by a coin cell battery and employing the all-in-one sensor BME280 (predecessor of current BME680) to acquire and advertise environmental data. In the results, the authors highlight the importance of performing an energy analysis in order to determine the correct operation of the device in terms of battery lifetime. Also, in \cite{dan2018}, the authors proposed a BLE beacon powered by a light EHU and a backup battery, capable of advertising air quality data every 60 seconds with an optimized power management strategy. Despite the advancements, these solutions still rely on batteries and broadcast the readings in a one-way connection to the gateways. Different from previous works, \cite{Fraternali2018} implemented a BLE device for indoor home applications that transmits data every 95 seconds and turns off when not operating. Furthermore, the work in~\cite{Sultania2022} designed a batteryless IoT node powered by an indoor solar panel that operates as a Low Power Node (LPN) node in a BLE mesh network, where the node establishes bi-directional communication with the gateway. The authors presented a comparison between different capacitor sizes and harvesting powers and concluded that the node shows a better performance with an optimal selection of the check interval for energy-aware communications. Although, it is valid to point out that BLE mesh may introduce delays in packet delivery due to data exchange between LPN-Friend nodes. Therefore, not being appropriate for applications that require high reliability. On this matter, \cite{10105285} demonstrated that is feasible to achieve highly reliable communications with BLE by implementing a low-latency architecture for BLE monitoring systems. 

On the other hand, in the context of LIoT communications, \cite{s21238024} proposed the design of a full-duplex IoT node powered by an indoor solar panel that uses organic Photovoltaic (PV) cells; with this, the authors demonstrated that sustainable components could be successfully integrated into IoT devices. In addition, our previous work \cite{Botirov2023} implemented a LIoT node in a SoC to transmit environmental sensing data, powered by an indoor light sustainable EHU. The results emphasized the relevance of employing SoCs given its many advantages, such as low-power operation and fast data acquisition.  

Despite the significant contribution highlighted in the previous works, the reliability in batteryless IoT applications, such as monitoring networks, still has ground to be improved. This motivates our work on reliable batteryless sensors with bi-directional communication employing BLE and LIoT.

\section{Components of a Batteryless IoT Sensor Node}
\label{Sec_III}

\begin{figure}[htbp]
\vspace{-0.5em}
\centering
\includegraphics[scale=0.14]{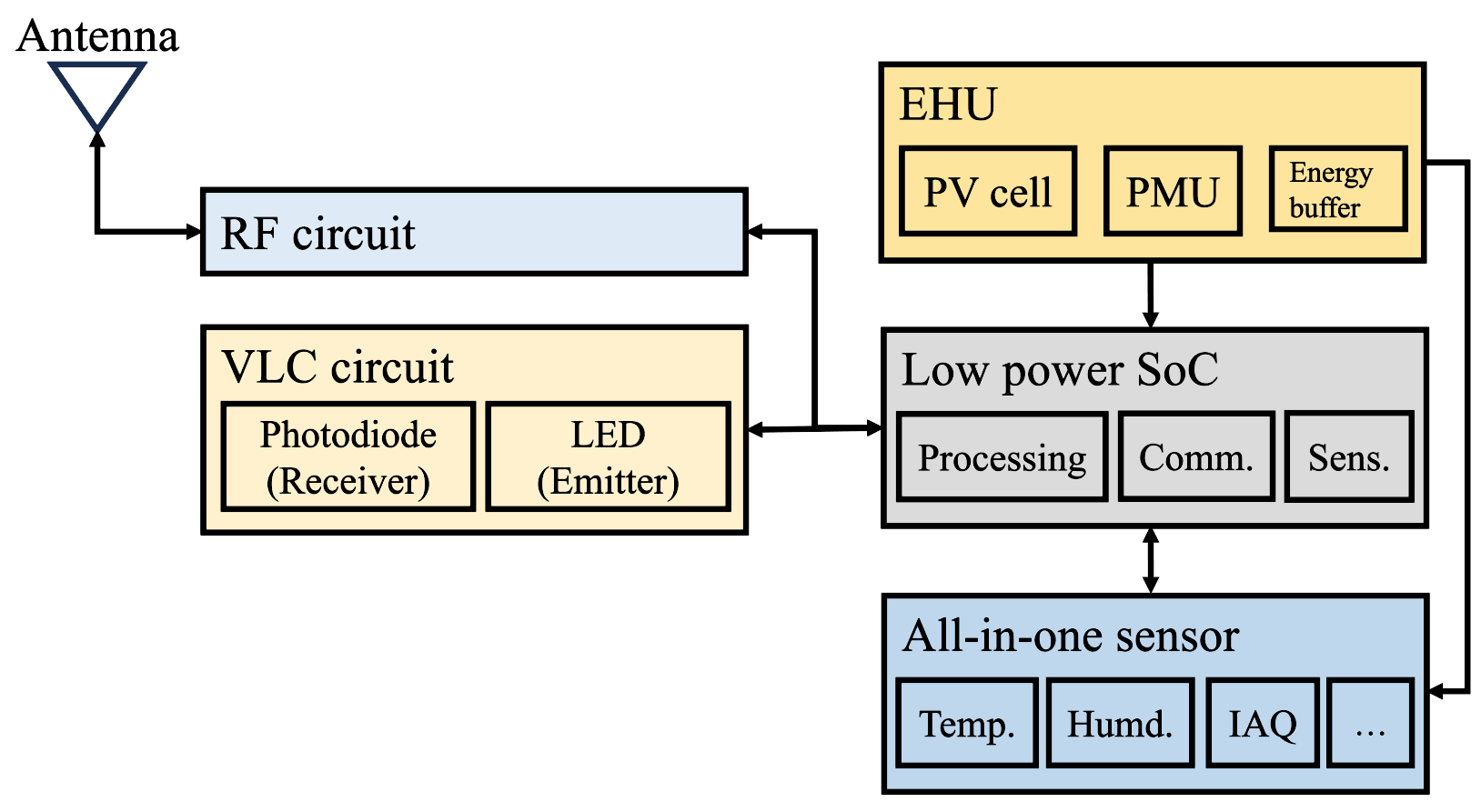}
\vspace{-0.5em}
\caption{Principal components of a batteryless IoT sensor node.}
\vspace{-1em}
\label{fig:net_diagram}
\end{figure}

The block diagram in Fig. \ref{fig:net_diagram} presents the principal components of a batteryless IoT device. The blocks illustrate the EHU, the different low-power sensors, the low-power SoC and integrated circuits, and the communication technologies. Next, we present a description of the main characteristics of each of these components, focusing on enhancing the reliability of the batteryless IoT sensor nodes.

\subsection{Energy Harvesting and Storage System}
One of the main components of the batteryless node design is the EHU, which produces and supplies the energy to the system. The first step to be considered is the selection of the EHU elements. Selecting organic PV cells and PE circuits can play an important role in offering a sustainable solution to harvest energy. Next, regarding efficient energy management, the Power Management Unit (PMU) in the EHU needs to allow the use of the Harvest-Storage-Use (HSU) approach to follow the ``Expose and Connect'' concept, allowing the EHU to power the system immediately after it harvests enough energy. This is possible with the Maximum Power Point Tracking (MPPT) approach, which maintains the optimum light energy harvesting point while transferring energy from the PV cell to a supercapacitor (SCap) as an energy buffer. Precisely, the use of SCaps instead of batteries is justified by several features, such as providing almost infinite charging cycles with high charge-discharge efficiency $(97\%-98\%)$, high power density, and no heat emission when draining \cite{2529}. 

\subsection{Low-Power "All-in-one" Environmental Sensors}
The condition of IAQ is closely related to the health state of indoor space occupants, along with the measurement of both temperature and humidity. For this reason, the selected sensor should be capable of measuring the principal components related to IAQ, such as temperature, humidity, air pressure, gas, and/or CO2 \cite{Santejudean2020}. Specifically, gas detection is regulated by the ISO16000-29 standard "Test methods for VOCs detectors." VOCs include Isoprene, Ethane, Acetone, Carbon Monoxide, and Ethanol, which are important to determining IAQ in laboratories and office buildings. However, all these parameters should be measured with the least energy consumption as possible. Thus, the sensor for indoor environmental sensing applications should always follow the state-of-the-art solution for ultra-low-power IoT applications.

\subsection{Low-power System-on-Chip}
A SoC refers to an integrated circuit that unifies a variety of electronic system peripherals, including memory, connections, and analog and digital components, all on a single substrate, which is controlled by a microprocessor. The SoC communication can be carried out through wired interfaces like I2C, SPI, UART, and wireless technologies such as WiFi or BLE. Hence, a SoC encapsulates all the necessary electronic components and circuits for sensing, processing, and communication. One of the most popular solutions is the ARM Cortex M series-based SoC, given its special design for applications demanding low-power consumption \cite{Botirov2023}.

\subsection{Reliable RF/Light Energy-Efficient Communication}

Wireless RF communication protocols like Bluetooth and BLE, ZigBee, and Wi-Fi are standard protocols for IoT sensor nodes. In particular, the BLE standard is the preferred protocol for low-power short-range IoT applications given its ultra-low-power characteristics when implemented in a SoC \cite{inproceedings1}. However, it is essential to exploit alternatives to complement radio communications and improve the reliability of batteryless IoT nodes. VLC is a prominent low-power choice to complement short-range radio communications since it covers the light spectrum from 380$\eta m$ to 780 $\eta m$, offering an alternative that improves the total reliability of indoor short-range networks.

\section{Energy Constraints for Continuous Batteryless Operation}
\label{Sec_IV}

The critical energy constraints when designing reliable batteryless IoT devices with continuous operation are threefold, energy harvesting efficiency, energy buffer capacity, and power consumption. Our design considers a constant illumination scenario to guarantee continuous operation and sensor readings at a fixed amount of time. Moreover, to achieve maximum sensing efficiency with the available energy, our solution proposes a power management strategy based on the periodic switching between active and sleep cycles.

The computation of the sleep and active cycles in the considered solution relies on the energy consumption analysis of the IoT nodes. This analysis employs the producer-consumer model for autonomous wireless devices presented in~\cite{Martinez2015}. The model considers all the energy sources and energy consumed by a wireless IoT node, and based on the energy conservation principle, outlines that the energy produced in a total operational time $T$ should be equal to or greater than the energy consumed, which is described in the following inequality: 
\begin{equation}
E_{buf}\!+\!\underbrace{\int_{0}^{T}\!P_{harv}(\tau)d\tau}_{E_{harv}(T)}\geq\underbrace{\int_{0}^{T}\!P_{dev}(\tau)d\tau}_{E_{dev}(T)},
\label{eq1}
\end{equation}
where $\tau$ is used as the variable of time, $E_{buf}$ represents the initial energy in the buffer in Joules, $P_{harv}(\tau)$ is the harvested power for an instant of time $\tau$ in Watts, and $P_{dev}(\tau)$ is the power consumed by the device for an instant of time $\tau$ in Watts. In addition, $E_{harv}(T)$ and $E_{dev}(T)$ represent the energy harvested and consumed in Joules during a complete duty cycle in $T$, respectively. $E_{buf}$ is the minimum amount of energy needed to ensure that the node stays operational, and our goal is not to go below it. Accordingly, a simplified version of the energy flow model is
\begin{equation}
E_{harv}(T)\geq\!E_{dev}(T), \forall T \in [0, \infty ).
\label{eq2}
\end{equation}
In other words, at each duty cycle, we aim to harvest more energy than what the node consumes for the active and sleep cycles. Therefore, the inequality described in \eqref{eq2} models the energy of a wireless IoT device that is always active. However, to show the energy produced during the sleep and active cycles in our low-power management implementation, we split the duty cycle time $T$ into the active and sleeping time, $T_a$ and $T_s$, being $T$ = $T_a$ + $T_s$. Consequently, we divided the energy consumed by the wireless node $E_{dev}(T)$ into the energy consumed during the active and sleeping time, $E_{deva}(T_a)$ and $E_{devs}(T_s)$, being $E_{dev}(T)= E_{deva}(T_a) + E_{devs}(T_s)$. Therefore, the energy conservation of a batteryless IoT node with low-power management is modeled as
\begin{equation}
E_{harv}(T_a+T_s)\geq\!E_{deva}(T_a)+E_{devs}(T_s).
\label{eq5}
\end{equation}
The obtained model is applied to design and implement the BLE and LIoT nodes. Specifically, it allows us to compute the correct sleeping period $T_s$ that satisfies \eqref{eq5}, enabling our proposed low-power management strategy for the batteryless IoT nodes.

\section{Design of Reliable IoT Sensor Nodes}
\label{Sec_V}
The design and implementation of the BLE and LIoT batteryless sensor nodes are described in \ref{BLEIoT} and \ref{LIoT}, respectively. Both subsections presented next are centered on three main parts: 1. Hardware and software; 2. Communication; 3. Low-power operation.

\subsection{BLE IoT Environmental Sensor Node}
\label{BLEIoT}
The batteryless BLE IoT node periodically senses and transmits environmental information such as VoCs for IAQ, air pressure, temperature, and humidity. After being acquired, those measurements are transmitted to the gateway in a bi-directional way, which indicates that during a connection, the gateway can also send instructions to the IoT node. 

\subsubsection{Hardware and software}
\begin{figure}[ht]
    \vspace{-0.5em}
    \centerline{\includegraphics[scale=.14]{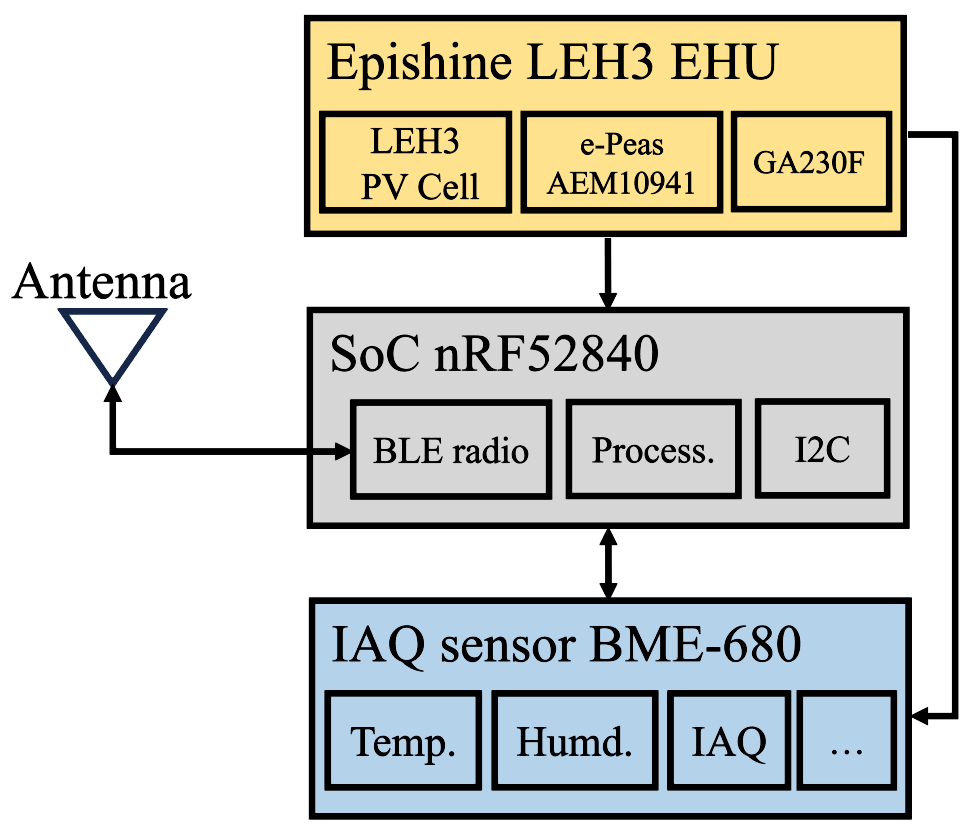}}
    \vspace{-0.5em}
    \caption{Hardware block structure of the BLE IoT node.}
    \vspace{-0.5em}
     \label{fig:BLE_structure}
\end{figure}

The hardware structure of the BLE IoT node, described in the blocks of Fig. \ref{fig:BLE_structure} is composed of the Epishine LEH3 EHU, the Bosch BME680 all-in-one sensor, and the Seeed XIAO BLE board with an ARM-based nRF52840 SoC, which enables Bluetooth~5.0 for wireless communication. The EHU consists of the e-Peas AEM10941 with the MPPT feature and of a 400mF SCap CAP-XX GA230F used as an energy buffer. A BME680 all-in-one low-power sensor is connected to the I2C port of the SoC board. Regarding the software, the communication and low-power algorithms are implemented over the manufacturer's firmware. These algorithms, based on open-source libraries, execute the sensor data acquisition and process the obtained information in data array structures, which are advertised as characteristics of the BLE Environmental Sensing Service~(ESS). The implemented processes are controlled by a timer-based scheduler that coordinates and controls the node sleep and active cycles.

\subsubsection{Communication}
\begin{figure}[ht]
    \vspace{-0.5em}
    \centerline{\includegraphics[scale=0.52]{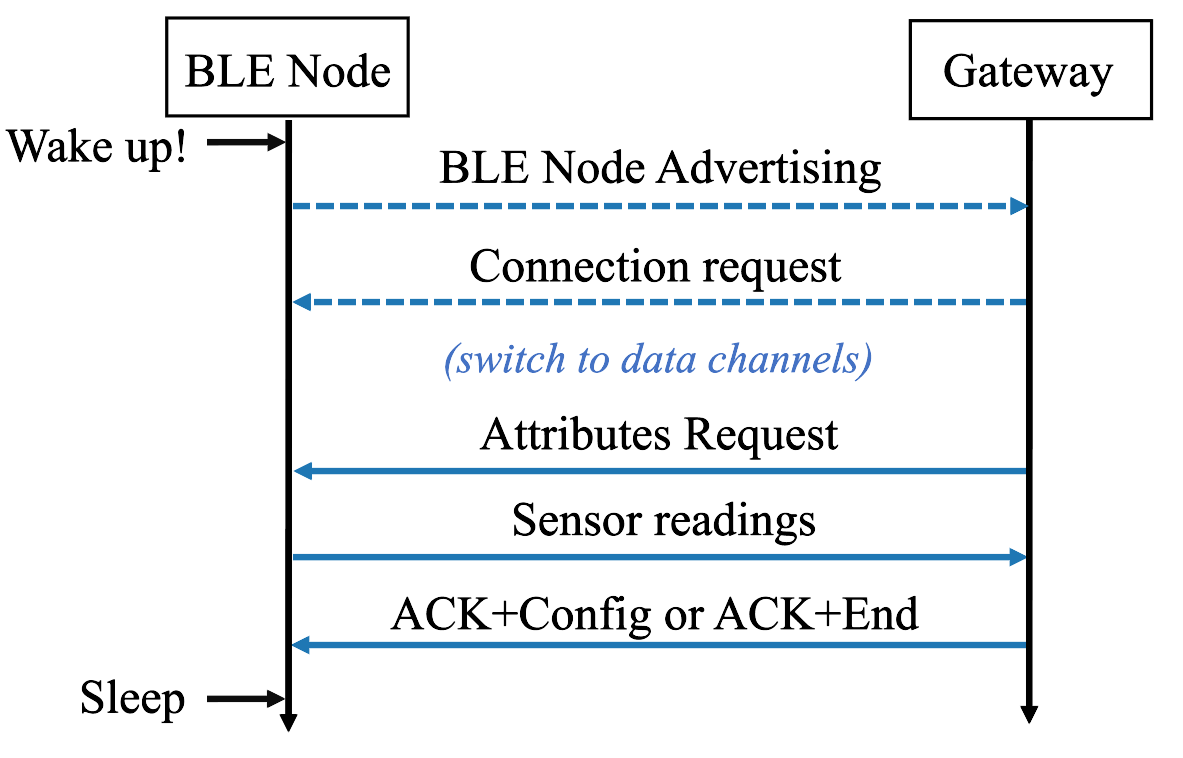}}
     \vspace{-0.5em}
    \caption{The BLE IoT node data exchange.}
    \vspace{-0.5em}
     \label{fig:BLE_dataexchange}
\end{figure}

The BLE IoT node enables low-power RF communication to the gateway through the Bluetooth radio chip incorporated in the SoC. The data exchange process, depicted in Fig. \ref{fig:BLE_dataexchange}, starts when the BLE node advertises the ESS service (Channels 37-39) for an interval of maximum~4s, with a transmission power of 4dBm. After the node's connection, the gateway requests the ESS attributes (Channels 0-36) that contain the sensor data. Finally, the node awaits a message from the gateway with configurations or to end the connection and go into the sleep cycle.

\subsubsection{Low-Power operation}
\begin{figure}[ht]
    \vspace{-0.5em}
    \centerline{\includegraphics[scale=0.132]{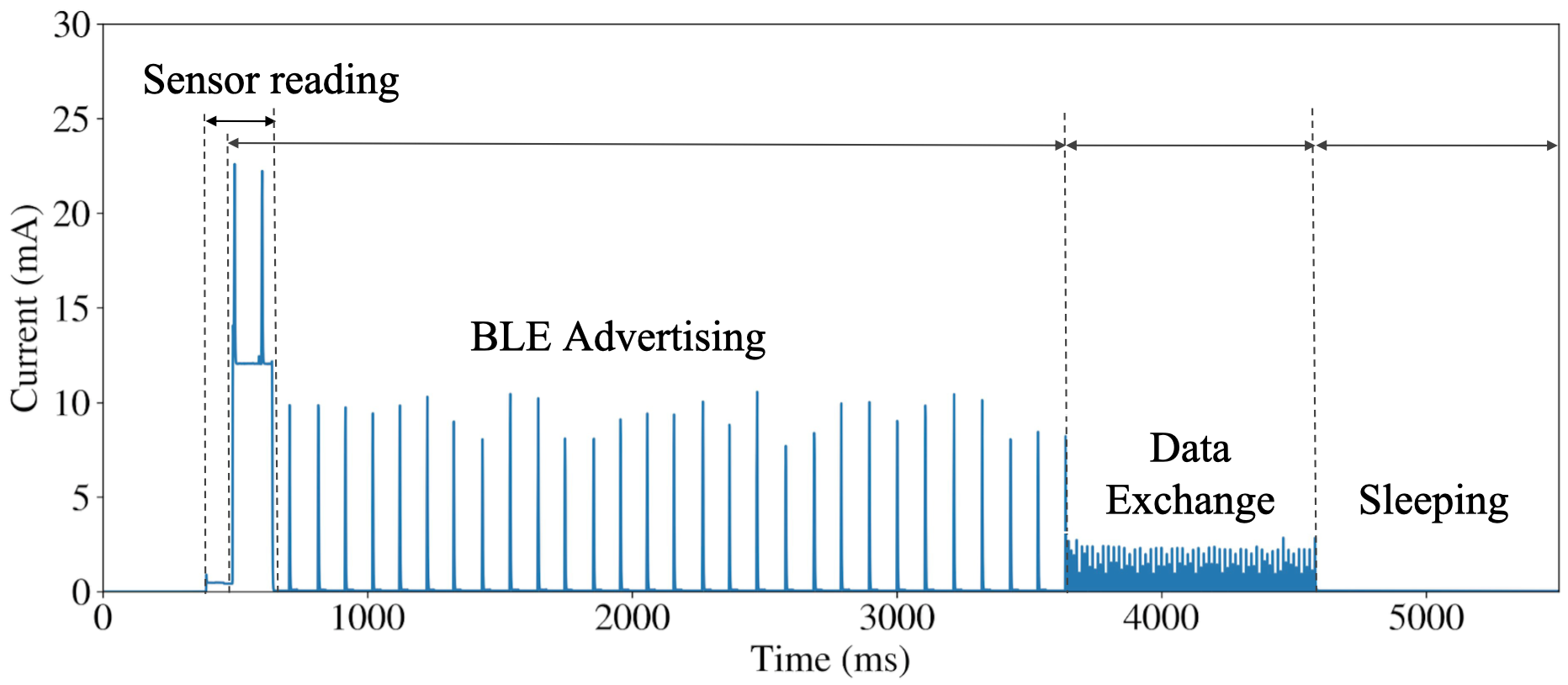}}
    \vspace{-1em}
    \caption{BLE IoT node energy consumption profile.}
    \vspace{-0.5em}
     \label{fig:all_sensorsBLE}
\end{figure}

To determine the optimal active and sleep cycles, we characterized the energy consumption of the IoT device by measuring the current consumed in every stage of the sensing and communication process. The current measurements for this analysis are carried out with the Nordic Semiconductors PPK II power profiler tool, which allows us to measure and log the current consumption at a maximum rate of $10^5$ samples per second. Accordingly, the different current consumption levels, in milliamperes, for every stage of the active and sleep cycles are shown in the energy consumption profile in Fig.\ref{fig:all_sensorsBLE}. Along with the current consumption, the figure also depicts the sleeping period, the sensor's data acquisition, and BLE advertising and data transmission. 

The detailed energy profile of the BLE IoT node allows the computation of the average current consumption for every stage of the active and sleep cycles. Consequently, under an operational voltage of $3.3 $V, and given the produced current of the EHU, $E_{harv}(T_a+T_s)$ can be  determined for two different illumination scenarios, with 500lx and 700lx. Next, with the performed current measurements, we obtain the values for the energy consumed by the devices in the active cycle $E_{deva}(T_a)$ and sleep cycle $E_{devs}(T_s)$. Finally, replacing the energy variables in~\eqref{eq5}, as described in subsection~\ref{Sec_IV}, the optimal value of the sleep cycle in seconds is determined, for both illumination scenarios, as $T_s$$=$$12.842$ and $T_s$$=$$20.520$, for 700lx and 500lx, respectively. These results enable the node to send up to 1491 samples from all sensors in a total operational period of 8 hours, considering an illumination of 700lx, and 1042 samples for 500lx. Table~\ref{tab:BLEenergy} summarizes the measured and computed current, time, and energy for every stage of the BLE IoT node in both illumination scenarios.

\begin{table}[h]
\vspace{-0.5em}
\caption{BLE IoT Batteryless Node Energy Consumption Profile}
\vspace{-0.5em}
\centering
\fontsize{7}{10}\selectfont 
\begin{tabular}{|cl|c|c|c|}
\hline 
\multicolumn{1}{|l|}{}                                                                                  & \textbf{Operation Stage} & \textbf{Current (mA)} & \textbf{Time (s)} & \textbf{Energy (J)} \\ \hline \hline
\multicolumn{1}{|c|}{\multirow{3}{*}{\textbf{\begin{tabular}[c]{@{}c@{}c@{}}Active \\ Cycle\\$E_{deva}$\end{tabular}}}} & Sensors Reading          & 7.550                     & 0.260                 & 0.0065                   \\ \cline{2-5} 
\multicolumn{1}{|c|}{}                                                                                  & BLE Advertising          & 0.400                     & 4.000                 & 0.0052                   \\ \cline{2-5} 
\multicolumn{1}{|c|}{}                                                                                  & Data Exchange            & 0.800                     & 1.300                 & 0.0034                   \\ \hline
\multicolumn{2}{|c|}{\textbf{Sleep Cycle $E_{devs}$ 700lx}}                                                                                         & 0.070                     & 12.842                 & 0.0029                   \\ \hline
\multicolumn{2}{|c|}{\textbf{Sleep Cycle $E_{devs}$ 500lx}}                                                                                         & 0.070                     & 20.520                 & 0.0047                   \\ \hline
\end{tabular}
\label{tab:BLEenergy}
\vspace{-1em}
\end{table}

\subsection{LIoT Environmental Sensor Node}
\label{LIoT}
Like the BLE node, the LIoT node also operates as an environmental sensor for indoor infrastructures. The main difference between them is the communication process. For the LIoT node, VLC is employed for downlink and IR for uplink, establishing bi-directional communication with the gateway.

\subsubsection{Hardware and software}
\begin{figure}[ht]
    \vspace{-0.5em}
    \centerline{\includegraphics[scale=0.14]{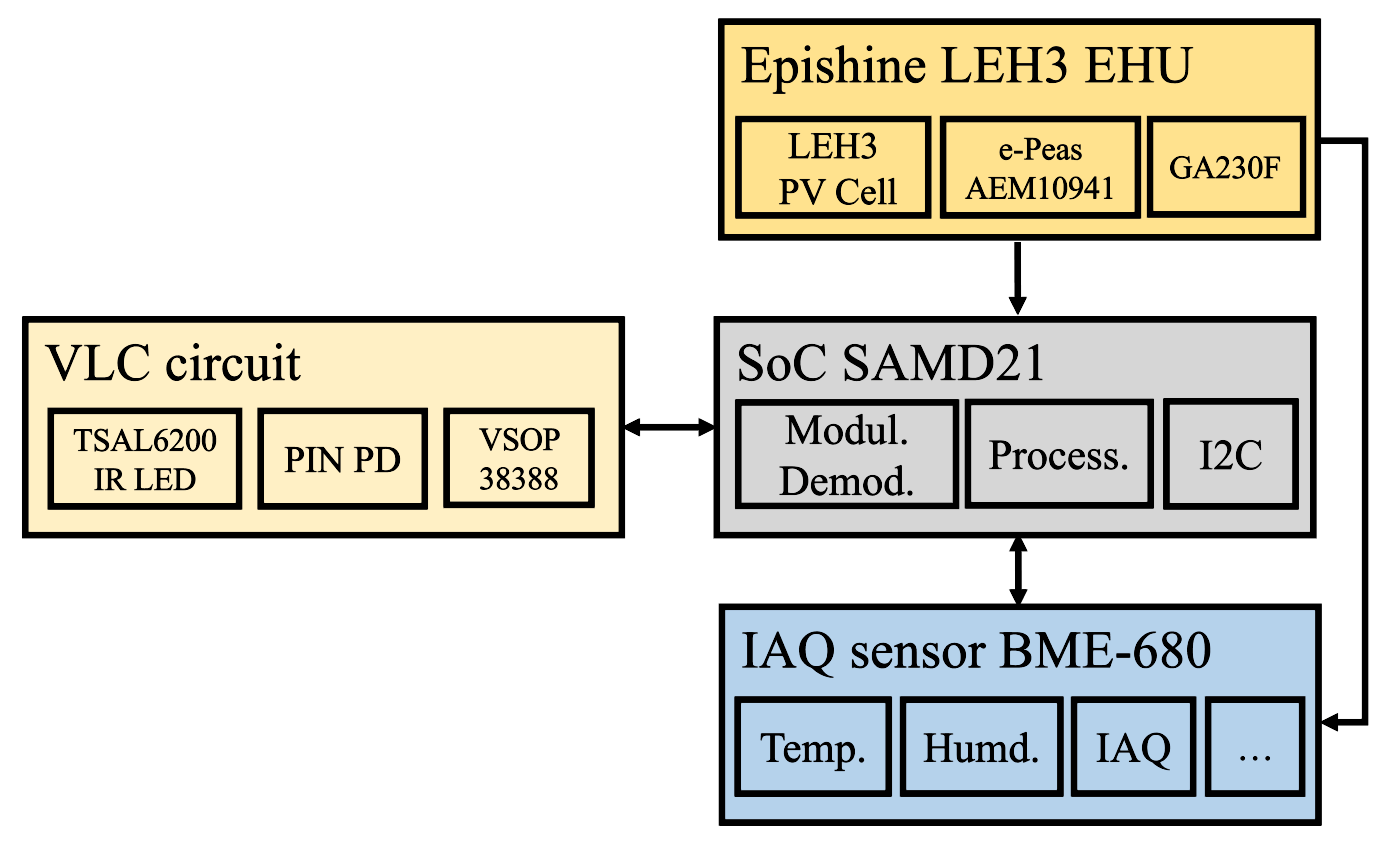}}
    \vspace{-0.5em}
    \caption{Hardware block structure of the LIoT node.}
    \vspace{-0.5em}
     \label{fig:LIoT_scheme}
\end{figure}

The first version of the LIoT node was introduced in detail in our previous work~\cite{Botirov2023}. For this current work, the LIoT node is also based on the Seed XIAO SAMD21 board and powered by the Epishine LEH3 EHU. Additionally, we incorporated the BME680 low-power sensor for sensing tasks since it adjusts perfectly with the evaluated use cases. The IoT node also has an additional transceiver unit for VLC/IR communication, and a Light-Dependent-Resistor (LDR) to determine the indoor illumination. Based on the obtained illumination, the node schedules the next sleep and active cycles. The LIoT transceiver unit comprises a photodiode operating at 525$\eta m$ to receive a downlink signal, an infrared light emitting diode TSAL6200 IRLED at 940$\eta m$ for uplink, and the all-in Integrated Circuit~(IC) VSOP38388 \cite{Botirov2023,s21238024}. The IC converts the current to voltage and amplifies it with an Automatic Gain Control~(AGC). Finally, a band-pass filter with a 38kHz center frequency filters the output signal. Both uplink and downlink signals are modulated at 38kHz carrier frequency with Pulse Position Modulation (PPM). Fig.~\ref{fig:LIoT_scheme} depicts the components for the sensing and communication features of the implemented LIoT node. Regarding the software, the implemented algorithms employ open-source libraries alongside the manufacturer board's firmware, handling the transmission modulation/demodulation and sensor readings through a timer-based scheduler.

\subsubsection{Communication}
\begin{figure}[ht]
    \vspace{-0.5em}
    \centerline{\includegraphics[scale=0.14]{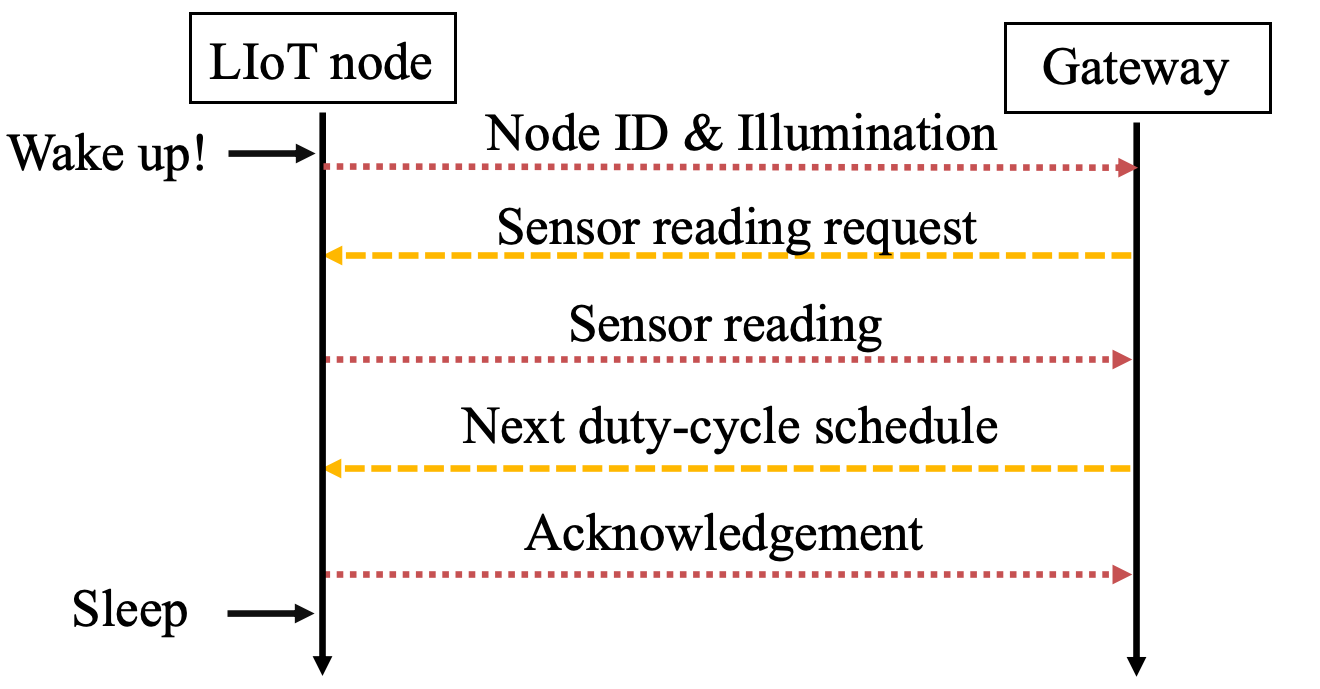}}
     \vspace{-0.5em}
    \caption{The LIoT node data exchange.}
    \vspace{-0.5em}
     \label{fig:LIoT_data_exchange}
\end{figure}

The communication process is carried out with the novel protocol depicted in Fig.\ref{fig:LIoT_data_exchange}. In our proposal, the LIoT node starts the data exchange by uploading the node ID and the illumination value over IR to the gateway. After that, the gateway requests the data from one or multiple sensors via VLC, and the LIoT node sends the requested readings. Finally, the gateway determines and sends the time value for the next sleep cycle, and receives the acknowledgment packet before closing the connection.

\subsubsection{Low-Power operation}
\begin{figure}[ht]
    \vspace{-0.5em}
    \centerline{\includegraphics[scale=.118]{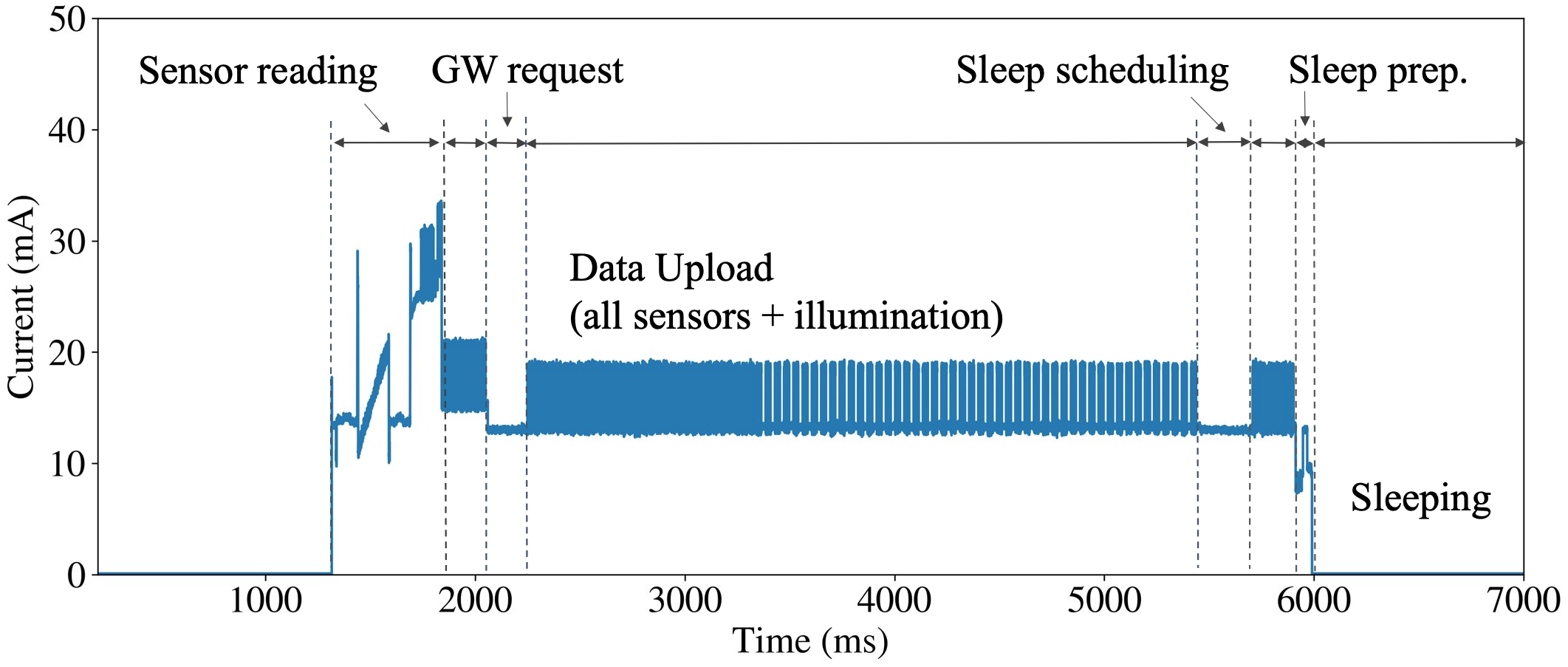}}
    \vspace{-0.5em}
    \caption{LIoT node energy consumption profile.}
     \label{fig:all_sensorss}
    \vspace{-0.5em}
\end{figure}

The energy consumption measurements are made in the same way as in the BLE IoT node. The active and sleep cycles are calculated using the measured average current when transmitting all sensor readings. Thus, following the process detailed in \ref{Sec_IV}, we determined the optimal sleep period in seconds for 700lx and 500lx scenarios, being $T_s$$=$$620$ and $T_s$$=$$1350$, respectively. These results enable the node to send up to 46 samples in a total operational time of 8 hours at 700lx, and 21 samples for 500lx. The average current consumption and times for the active and sleep cycles (when all the sensor readings are requested) are detailed in Table~\ref{tab:LIoTenergy}. In addition, Fig. \ref{fig:all_sensorss} shows the current consumed in the different sensing/communication stages of the node. Compared to~\cite{Botirov2023}, we can observe that the active periods are longer due to the amount of transmitted data.

\begin{table}[h]
\vspace{-1em}
\caption{LIoT Batteryless Node Energy Consumption Profile}
\vspace{-0.5em}
\centering
\fontsize{7}{10}\selectfont 
\begin{tabular}{|cl|c|c|c|}
\hline
\multicolumn{1}{|l|}{}                                                                                  & \textbf{Operation Stage} & \textbf{Current (mA)} & \textbf{Time (s)} & \textbf{Energy (J)} \\ \hline \hline
\multicolumn{1}{|c|}{\multirow{4}{*}{\textbf{\begin{tabular}[c]{@{}c@{}}Active \\ Cycle\end{tabular}}}} & GW Request               & 12.69                     & 0.428                 & 0.0179                   \\ \cline{2-5} 
\multicolumn{1}{|c|}{}                                                                                  & Sensor Reading           & 17.73                    & 0.525                 & 0.0307                   \\ \cline{2-5} 
\multicolumn{1}{|c|}{}                                                                                  & Data Upload              & 14.58                    & 3.58                 & 0.1722                   \\ \cline{2-5} 
\multicolumn{1}{|c|}{}                                                                                  & Sleep Set                & 9.81                     & 0.078                 & 0.0025                   \\ \hline
\multicolumn{2}{|c|}{\textbf{Sleep Cycle $E_{devs}$ 700lx}}                                                                                         & 0.087                     & 620                 & 0.1780                   \\ \hline
\multicolumn{2}{|c|}{\textbf{Sleep Cycle $E_{devs}$ 500lx}}                                                                                         & 0.087                     & 1350                 & 0.3876                   \\ \hline
\end{tabular}
\label{tab:LIoTenergy}
\vspace{-1em}
\end{table}

\section{Experimental Performance Evaluation}
\label{Sec_VI}

In order to validate the implemented BLE and LIoT nodes, and to test the system stability, we conducted two experiments of eight hours each. We consider an office environment with two levels of constant illumination, 700lx~$\pm$5\% and 500lx~$\pm$5\%. In this scenario, the gateway is placed half a meter from the LIoT node and three meters from the BLE IoT node. During the experiments, the nodes' current consumption and the SCap's charge-discharge voltage are measured to show the energy conservation during active and sleep cycles. 

In particular, for the BLE node, an additional 5\% of time is included in the duty cycle to compensate for temperature and humidity changes in the environment. Finally, assuming a shorter BLE advertising time and small changes in energy consumption in practice, the timer-based scheduler of the BLE node starts a new duty cycle every 19.3 and 26.7 seconds for 700lx and 500lx, respectively.

\begin{figure}[ht]
    \vspace{-0.5em}
    \centerline{\includegraphics[scale=.453]{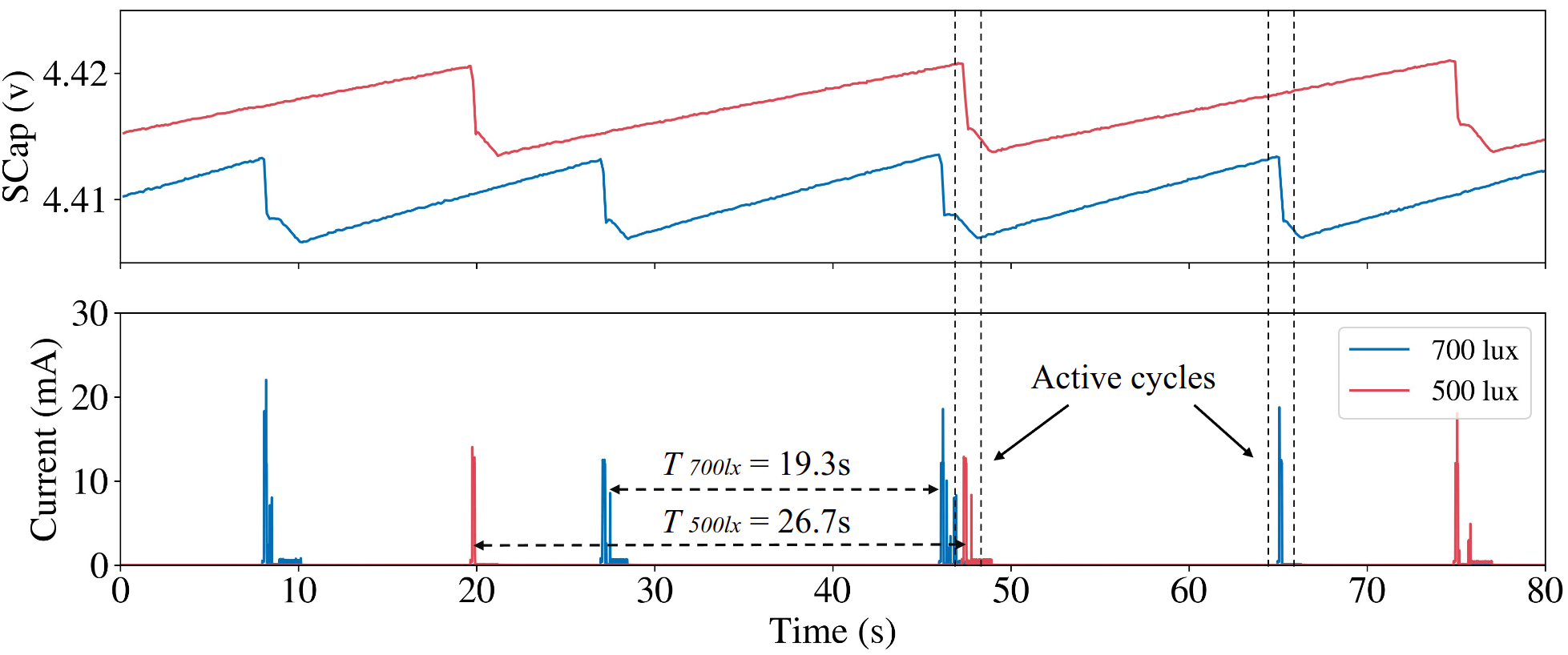}}
    \vspace{-0.5em}
    \caption{Energy conservation in the BLE IoT node.}
     \label{fig:current_voltage_ble}
    \vspace{-0.5em}
\end{figure}

Fig. \ref{fig:current_voltage_ble} shows the current consumed by the BLE IoT node (below) and the voltage charge-discharge of the SCap (above) versus time for 700 and 500lx. Note that for both illumination scenarios, the SCap experiments a discharge of approximately 0.007V during the active cycle and recovers during the sleep cycle, highlighting the energy conservation principle for this implementation. Also, we can observe that the recovery time is longer for the 500lx illumination scenario, which is expected given the lower amount of energy harvested. 

\begin{figure}[ht]
    \vspace{-0.5em}
    \centerline{\includegraphics[scale=.438]{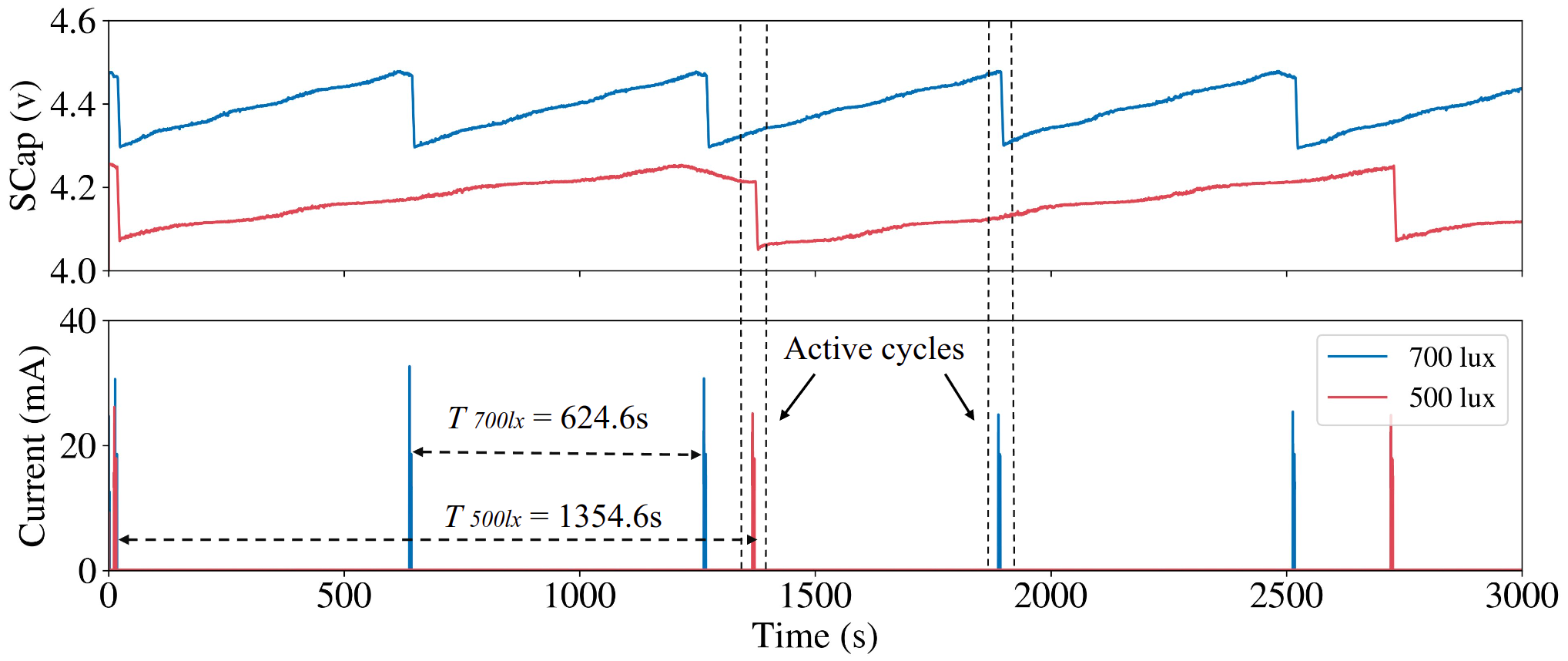}}
    \vspace{-0.5em}
    \caption{Energy conservation in the LIoT node.}
     \label{fig:current_voltage_vlc}
     \vspace{-0.5em}
\end{figure}

The energy conservation in the LIoT node can be observed in Fig. \ref{fig:current_voltage_vlc}. The LIoT node can perform a complete sensor measurement every 624.6s at 700lx and 1354.6s at 500lx. Compared to the BLE IoT node, the measurements show a longer duty cycle and higher current consumption, leading to a 0.17V discharge on the SCap. This is because of the energy consumed by the external transceivers used for transmission via IR and VLC. However, the performance of the LIoT node satisfies the necessities for environmental sensing, establishing continuous communication with the gateway and complementing the RF transmission of the BLE IoT node.

Table~\ref{tab:8hourtest} summarizes the results obtained in the performed experiments.
On that, PDR stands for packet delivery rate, which corresponds to the ratio of the successful duty cycles to the failed ones. In the case of BLE, a packet is successfully delivered when the gateway receives the ESS attributes and closes a connection, and for the LIoT, a packet is successfully delivered when all the sensor readings are received and decoded by the gateway. The BLE node achieved a PDR of 0.991 and 0.912 for the 700 and 500lx scenarios, respectively. On the other hand, the LIoT node successfully delivered all the sent packets, having a PDR of 1.000 for both scenarios. 

\begin{table}[ht]
\caption{Batteryless IoT Nodes' Performance Evaluation (8 hours)}
\centering
\fontsize{7}{10}\selectfont 
\begin{tabular}{|c|c|c|c|c|l|}
\hline
\textbf{Node}                     & \textbf{Illumination} & \textbf{\begin{tabular}[c]{@{}c@{}}Packets\\ Sent\end{tabular}} & \textbf{\begin{tabular}[c]{@{}c@{}}Packets\\ Received\end{tabular}} & \textbf{PDR} & \textbf{\begin{tabular}[c]{@{}l@{}}Avg.\\ SCap (V)\\\end{tabular}} \\ \hline\hline
\multirow{2}{*}{\textbf{BLE IoT}} & 700lx               & 1491                                                            & 1479                                                                & 0.991        & 4.463                                                              \\ \cline{2-6} 
                                  & 500lx               & 1042                                                               & 951                                                                   & 0.912            & 4.416                                                                   \\ \hline
\multirow{2}{*}{\textbf{LIoT}}    & 700lx               & 46                                                              & 46                                                                  & 1.000        & 4.235                                                               \\ \cline{2-6} 
                                  & 500lx               & 21                                                               & 21                                                                   & 1.000            & 4.353                                                                   \\ \hline
\end{tabular}
\label{tab:8hourtest}
\vspace{-1em}
\end{table}

\section{Conclusions and future work}
\label{Sec_VIII}
In this work, we explored the design and implementation of BLE and LIoT batteryless devices powered by an indoor light EHU. We analyzed the energy consumption of the nodes and calculated the optimal sleep period for the operation based on the energy harvested. The obtained results demonstrate the feasibility of having an energy-autonomous operation of our proposed IoT node designs under environments with two different light intensities. In particular, our design guarantees the transmission of one packet (at 700lx) every 19.3 and 624.6 seconds for the BLE and LIoT nodes, respectively, satisfying the measurement frequency requirements for an environmental-sensing use case. Moreover, our measurements showed that more than 90\% of the packets were successfully delivered for BLE and LIoT. Future research directions include testing our solution in different scenarios and exploiting the capacities of the SoC to process the acquired data for intelligent edge decision-making.
\bibliographystyle{ieeetr}
\bibliography{bibliography}

\end{document}